\documentclass[aps,pre,preprint,groupedaddress,showpacs]{revtex4}
\usepackage{epsfig,bm,latexsym,amssymb,amsfonts,amsmath,graphicx}
\usepackage{amsmath}
\usepackage{amsfonts}
\usepackage{epsf}
\usepackage{natbib}
\def\beq{\begin{equation}}
\def\eeq{\end{equation}}

\newcommand{\mbf}{\mathbf}

\usepackage[usenames]{color}
\definecolor{Red}{rgb}{1,0,0}

\begin{document}
\title{Variational method for estimating the rate of convergence
of Markov Chain Monte Carlo algorithms}

\author{Fergal P. Casey}
\email[]{fergal.casey@ucd.ie}
\affiliation{Complex and Adaptive Systems Laboratory, University College Dublin, Dublin~4, Ireland}
\altaffiliation[Also at ]{UCD Conway Institute of Biomolecular and Biomedical Research, University College Dublin, Dublin~4, Ireland}
\author{Joshua J. Waterfall}
\affiliation{Department of Molecular Biology and Genetics, Cornell University, Ithaca, NY 14853, USA}
\author{Ryan N. Gutenkunst}
\affiliation{Department of Biological Statistics and Computational Biology, Cornell University, Ithaca, NY 14853, USA}
\author{Christopher R. Myers}
\affiliation{Computational Biology Service Unit, Life Sciences Core Laboratories Center, Cornell University, Ithaca, NY 14853, USA}
\author{James P. Sethna}
\affiliation{Laboratory of Atomic and Solid State Physics, Cornell University, Ithaca, NY 14853, USA}

\date{\today}

\begin{abstract}
We demonstrate the use of a variational method to determine a quantitative lower 
bound on the rate of convergence of Markov Chain Monte Carlo (MCMC) algorithms as
a function of the target density and proposal density. 
The bound relies on approximating the second largest eigenvalue in the spectrum
of the MCMC operator using a variational principle and the approach is applicable
to problems with continuous state spaces. 
We apply the method to one dimensional examples with Gaussian and
quartic target densities, and we contrast the performance of the 
Random Walk Metropolis-Hastings (RWMH) algorithm with a ``smart'' variant 
that incorporates gradient information into the trial moves, a generalization
of the Metropolis Adjusted Langevin Algorithm (MALA). 
We find that the variational method agrees quite closely with numerical 
simulations. We also see that the smart MCMC algorithm often fails to 
converge geometrically in the tails of the target density except in 
the simplest case we examine, and even then care must be taken to choose 
the appropriate scaling of the deterministic and random parts of the
proposed moves. Again, this calls into question the utility of smart MCMC in more
complex problems.
Finally, we apply the same method to approximate the rate of convergence in 
multidimensional Gaussian problems with and without importance sampling.
There we demonstrate the necessity of importance sampling for target densities 
which depend on variables with a wide range of scales.

\end{abstract}

\maketitle

\section{Introduction}
Markov Chain Monte Carlo (MCMC) methods are important tools in parametric 
modeling~\citep{mcmcinpractice,tarantola} where the goal is to determine a 
posterior distribution of parameters given a particular dataset. 
Since these algorithms tend to be computationally
intensive, the challenge is to produce algorithms that have better convergence
rates and are therefore more efficient~\citep{atchade:0305,bedard2:0106}. Of particular
concern are situations where there is a large range of scales associated with the target
density, which we find are widespread in models from many different fields~\citep{kevin1,kevin2,jacobsen, Josh, Ryan}.

In this manuscript we quantify the convergence of the MCMC method by
the second largest eigenvalue in absolute value for the associated operator 
in $L^2$. This is not the only numerical quantity that can be used to 
describe the convergence properties. Other authors quantify convergence
with different metrics: computing the constant of geometric convergence with 
respect to the total variation norm~\citep{computablebounds}, monitoring sample 
averages~\citep{brooksroberts}, evaluating mixing of parallel chains~\citep{gelmanrubin}
or looking at the integrated autocorrelation time of functions of the 
sample~\citep{optimalscaling,weakconvergence}. The connection between the 
second eigenvalue and total variation norm is discussed in~\citep{L2convergence}.
To connect the second eigenvalue estimates to metrics based on autocorrelation,
we would argue informally that the second eigenvalue determines the autocorrelation
time of the slowest mixing function of the sample and as such represents a 
``worst'' case for the length of time you would need to run the chain to reduce 
the variance of sample averages to a predefined level.  

There are a number of techniques to either determine exactly or bound the 
second eigenvalue or the constant of geometric convergence for MCMC algorithms 
on discrete state spaces~\citep{rapidmixing,frigessi,jerrumsinclair,diaconis}, 
but the methods for finding quantitative bounds for continuous state spaces 
require a more technical formulation. 
Where work has been done in that area,
upper bounds on the convergence rate can be derived using purely 
analytical~\citep{rosenthal,joneshobert,computablebounds} or semi-analytical 
techniques~\citep{garren}, but may not always be very useful for selecting
parameters optimally. 
Therefore, in this work, we show that a conceptually straightforward
variational method can provide convergence rate estimates for continuous 
state space applications. In contrast to earlier closely related 
work~\citep{L2convergence,optimalscaling}, we move away from mathematical 
formalities, focussing from the start on specific examples and step through the 
calculations that provide the second eigenvalue bounds. In~\citet{optimalscaling},
rules of thumb are provided for determining the optimal acceptance rate and
step lengths for both the Random Walk Metropolis-Hastings and the Metropolis 
Adjusted Langevin Algorithm, in the asymptotic limit of infinite dimensions
where it can be proved that those methods are approximated by diffusion
processes. The rules are widely used as they are independent of the specific
form of the target density, appear from numerical simulations to be
appropriate far from the infinite dimensional asymptotic limit and are easily
implemented. In contrast, the approach proposed here is to establish 
convergence properties for particular MCMC algorithms based on their performance
on simple target distributions without the need to set up a diffusion approximation
in an infinite dimensional limit. Poor performance or lack of convergence
on these simple distributions then indicates that further application with
more complex target densities will also suffer from convergence problems. 
Conversely, an identification of a range of parameters which provide good convergence
properties for simple target distributions may be used as a starting point for
further applications.
Even though we provide only lower bounds on the
second eigenvalue we show these bounds can be remarkably tight due to careful
choice of test functions, and computing the approximate convergence rate as 
a function of algorithm parameters allows us to optimally tune those parameters.

We have been able to obtain explicit formulas for one dimensional example 
problems but the method may be more generally applicable, when applied in an 
approximate way, as we demonstrate for a multidimensional problem.
\section{Markov Chain Monte Carlo}
Typically, one wishes to obtain a sample $x_1,x_2...$ from a probability 
distribution $\pi(x)$ which is sometimes called the \emph{target} distribution. 
An MCMC algorithm works by creating a Markov chain that has $\pi(x)$ as its
unique stationary distribution, i.e. after many steps of the chain any initial 
distribution converges to $\pi(x)$. 
A sufficient condition to establish $\pi(x)$ as the stationary distribution
is that the chain be ergodic and that the 
\emph{transition density}, $t(x,y)$, of the chain satisfy detailed balance: 
\[ \pi(x) \, t(x,y) = \pi(y) \, t(y,x) .\]
Given a \emph{proposal density} $q(x,y)$ for generating 
moves, one way to construct the required transition density~\citep{casella,metropolis} 
is to define
$t(x,y) = \alpha(x,y) \, q(x,y)$ where
\begin{equation} 
\alpha(x,y) = \mbox{min}\left(\frac{q(y,x)\pi(y)}{q(x,y)\pi(x)},1\right) 
\label{acceptanceratio}
\end{equation} 
is the \emph{acceptance probability} of the step $x \to y$. 
Obtaining the sample from the stationary
distribution then involves letting the chain run past the transient (\emph{burn-in})
time and recording iterates from the late time trajectory at time intervals
exceeding the correlation time. How long
it takes to reach the stationary distribution determines the efficiency of the 
algorithm and for a given target distribution, clearly it depends on the choice
of the proposal density.
We can write down the one-step evolution of a probability density $p(x)$
as a linear operator: 
\begin{eqnarray*} 
(\mathcal{L}p)(y) &=&\int t(x,y)p(x)\,dx + 
\left( 1 - \int t(y,x)\,dx \right)p(y) \\ 
&=&\int \left( t(x,y)p(x) - t(y,x)p(y) \right) \,dx + p(y)  
\end{eqnarray*} 
where $dx = dx_1 \ldots dx_n$, $dy = dy_1 \ldots dy_n$, $n$ is the dimension of the state space and
all integrals are from $-\infty$ to $\infty$ here and elsewhere in this manuscript.
The second form makes it explicit that $p(y) = \pi(y)$ is the stationary distribution
by the detailed balance relation.

Now, if the linear operator has a discrete set of eigenfunctions and eigenvalues, 
it holds that the asymptotic convergence rate is determined by the second largest
eigenvalue in absolute value (the largest being one)~\citep{lawler,lambda2convergence}. We
will write this eigenvalue as $\lambda^*$, and will refer to it as the \emph{second eigenvalue} 
meaning the second largest in \emph{absolute} value.
Geometric convergence of the chain is ensured when $\lambda^*<1$~\citep{L2convergence}, and then
the discrepancy between the density at the $m^{th}$ iterate of the chain and the target density 
decreases as $(\lambda^*)^m$ for large $m$. Many previous authors have taken this second 
eigenvalue approach, in both the finite and continuous state space settings~\citep{diaconis,
frigessi,rosenthal2,jerrumsinclair,garren}, as it provides a useful quantifier for the convergence
rate. Ideally we would like algorithm parameters to be adjusted such that $\lambda^*$ is as 
as small as possible.

The variational calculation allows us to obtain an estimate for $\lambda^*$, 
but before we can do this we need to convert our operator into a 
self-adjoint form which ensures that the eigenfunctions are orthogonal. This 
is easily accomplished by a standard technique~\citep{rapidmixing} of 
modifying the transition density by 
$s(x,y) = t(x,y) \sqrt{\pi(x)}/\sqrt{\pi(y)}$ 
and our self-adjoint operator is then given by
\begin{eqnarray} 
(\mathcal{S}p)(y) &=&\int s(x,y)p(x)\,dx + 
\left(1 - \int t(y,x)\,dx \right) p(y) \\ 
&=&\int \left( s(x,y)p(x) - t(y,x)p(y) \right) \,dx + p(y)  
\label{Soperator}
\end{eqnarray} 
where the ``diagonal'' part of the old operator (multiplying $p(y)$) need not be transformed 
using $s(x,y)$. It is easy to show that defined as above, $\mathcal{S}$ is self-adjoint using 
the standard inner product in $L^2$ with respect to Lebesgue measure.
Note that if $u(x)$ is an eigenfunction of the operator $\mathcal{S}$, then 
$\sqrt{\pi(x)}u(x)$ is an eigenfunction of the original operator $\mathcal{L}$ with the
\emph{same} eigenvalue.
\subsection{Metropolis-Hastings and smart Monte Carlo} 
We consider two MCMC algorithms which essentially differ only in the choice of 
proposal density and acceptance probability that is used in selecting steps. 
We will refer to the \emph{Random Walk Metropolis-Hastings} (RWMH) algorithm
as that which uses a symmetric proposal density to determine the next move; for 
example, a Gaussian centered at the current point: \\ 
$q(x,y) = \sqrt{|L|/(2 \pi)} \exp 
\left( - (y-x)^T L (y-x)/2 \right) $
where $L$ is an inverse covariance matrix that needs to be chosen appropriately
for the given problem (\emph{importance sampling}). In other words, the proposed
move from $x$ to $y$ is given by $y = x + R$ where $R \sim \mbf{N}(0,L^{-1})$ is
a normal random variable, mean $0$ and covariance $L^{-1}$. Thus the update on the
current state has no deterministic component. We will see that when the target density is not
spherically symmetric, a naive implementation of the Metropolis-Hastings algorithm 
where the step scales are all chosen to be equal leads to very poor performance of 
the algorithm. As would be expected the convergence deteriorates as a function of 
the ratio of the true scales of the target density to the scale chosen for 
the proposal density.

One variant used to accelerate the standard algorithm is a \emph{smart} 
Monte Carlo method~\citep{smartmc} that uses  
the gradient of the negative of the log target density at every step, 
$G(x) = -\nabla \log(\pi(x))$ to give 
\begin{equation}
q(x,y) = \frac{\sqrt{|L|}}{\sqrt{2\pi}} 
\exp\left(-\frac{1}{2} (y-(x-H^{-1} G(x))^T L (y-(x-H^{-1} G(x))\right)
\label{smcequation}
\end{equation}
and $H$ can be considered either as a constant scaling of the gradient part of 
the step or, if it is the Hessian of $-\log(\pi(x))$, 
as producing a Newton-like optimization step~\citep{trustregion}. The move to $y$ is generated as 
$y = x - H^{-1} G(x) + R$, so now we have a \emph{random} component 
$R \sim \mbf{N}(0,L^{-1})$ and a \emph{deterministic} component
$-H^{-1} G(x)$. Viewed like this, moves can be considered to be
steps in an optimization algorithm (moving to maximize the probability of 
the target density) with random noise added. We will see that with an optimal
choice of $H$ and for Gaussian target densities, the smart Monte Carlo method 
can converge in one step to the stationary distribution. We will also see 
that for a one dimensional non-Gaussian distribution it actually fails to converge, 
independent of the values of the scaling parameters.
\subsection{Variational method}
Once we have the self-adjoint operator for the chain, 
$\mathcal{S}$ from Eqn. \ref{Soperator},  and we know 
the eigenfunction with eigenvalue $\lambda_1 = 1$, $\sqrt{\pi(x)}$, we can 
look for a candidate second eigenfunction in the function space orthogonal to 
the first eigenfunction where the inner product is defined by 
$(p_1,p_2) = \int p_1(x)p_2(x) \,dx $. 
Given a family of normalized candidate functions in this space, 
$v_a(x)$, with variational parameter $a$, the variational 
principle~\citep{variationalbound,lawler} states
\begin{equation}
\mbox{max}_a |(v_a,\mathcal{S}v_a)| \le \lambda^* \le 1
\label{varbound}
\end{equation}
and depending on how accurately our family of candidate functions captures the
true second eigenfunction, this can give quite a close approximation to the 
second dominant eigenvalue. In the problems we examine in the following sections
the target densities have an even symmetry which makes it straightforward to 
select a variational trial function: any function with odd symmetry will naturally
lie in the orthogonal space. For more complicated problems with known symmetries this 
general principle may be useful in selecting variational families for the purposes
of algorithm comparison. Another approach to constructing the variational family is shown 
in the section on multidimensional target densities: choose the test function as a linear
combination of two functions, one with the properties that are required 
(i.e. slow convergence to the target distribution) and then the additional term is used
merely to preserve orthogonality.


This variational method for providing a lower bound to the second eigenvalue of the MCMC
algorithm was foreshadowed by a similar approach of Lawler and Sokal~\citep{lawler}. These authors
considered the flow of probability out of
a subset A of the state space; in our language, their test functions were confined to the 
family
	$v_A(x) = (\pi(A^c) \chi_A - \pi(A) \chi_{A^c})/(\pi(A) \pi(A^c))$, where $\chi_{A}$ is the
indicator function on the set $A$. By allowing for more general test functions, we establish 
not only rigorous but also relatively tight bounds on convergence rates, providing guidance for 
parameter optimization and algorithm comparisons.

Writing out explicitly for $\mathcal{S}$ in $(v_a,\mathcal{S}v_a)$ we have
\begin{equation}
(v_a,\mathcal{S}v_a) = \int \int v_a(x) s(x,y) v_a(y) \, dx dy -
\int \int t(y,x) \left( v_a(y) \right)^2 \, dx dy + 1 \enspace .
\label{scalarprod}
\end{equation}
As we will 
see in the following section, the lower bound in Eqn. \ref{varbound} can be 
arbitrarily close to $1$ and therefore equality holds. In these situations
we can also show that the chain does not converge geometrically, based on the total variation
norm definition of geometric convergence~\citep{computablebounds}. 
However, whether the type of convergence changes or not, we still refer to the magnitude of 
the second eigenvalue estimate in determining efficiency of the algorithm. The 
rationale is that the second eigenvalue determines the longest possible autocorrelation time of
a function of the MCMC sample; the worst case autocorrelation time will be of the order 
$1.0/log(\lambda^*)$ which could be extremely long.
We will also see that there can be eigenvalues in the spectrum that are close 
to $-1$ which determine the asymptotic convergence rate, i.e. $\lambda^* = |\lambda_n|$ where $\lambda_n < 0$. 
Interestingly, for this situation there is oscillatory behavior of the Markov 
chain. 
\section{Examples}
\subsection{Gaussian target density}
Consider the simplest case of a one dimensional Gaussian target density
$\pi(x) = \sqrt{k/(2\pi)} \exp(-k x^2/2) $ with variance
$1/k$. Under the RWMH algorithm, the proposal 
density is 
\begin{equation}
q(x,y) = \sqrt{\frac{l}{2\pi}} \exp 
\left(-\frac{1}{2} l (y-x)^2 \right) \enspace .
\label{1dqxy}
\end{equation}
The issue is to determine $l$ optimally; a first guess would be that
$l = k$ is the best choice. The rationale behind this is that since the target and 
proposal densities have the same form, if they also have the same scales, then the convergence 
rate might be expected to be optimal. 
We will see that this is not actually correct. 

To begin, define a variational function $v_a(x) \propto x \exp(-a x^2/2)$, 
orthogonal to the target density and normalized such
that $\int v_a^2 \, dx = 1$. We can motivate this choice by recognizing
that any initial distribution that is asymmetric will most likely
have a component of this test function, and a convergence rate 
estimate based on it roughly corresponds to how fast probability
``equilibrates'' between the tails. More commonly, variational calculations
will use linear combinations of many basis functions with
the coefficients as variational parameters. We find here that including higher
order terms in the test function is unnecessary as we obtain tight enough 
bounds just retaining the lowest order term.

We proceed by evaluating Eqn. 
\ref{scalarprod} noting that because of the form of the acceptance
probability, Eqn. \ref{acceptanceratio}, there are two functional
forms for the kernels $t(x,y)$ and $s(x,y)$ depending on the sign of
$y^2 - x^2$, {i.e.} whether the ``energy'' change, 
$\Delta E(x,y) = -\log(\pi(y)) + \log(\pi(x)) = k (y^2-x^2)/2$,   
is positive or negative. It is then convenient to use the coordinate transformation 
$y = rx, x = x$ or $x = ry, y = y$ where $-1\le r \le 1$ and $-\infty \le x,y \le \infty$ to 
evaluate the integrals. An explicit 
expression for $(v_a,\mathcal{S}v_a)$ can be obtained for this case of 
a Gaussian target density. 

\begin{figure}[hbtp]
		\begin{center}
		\epsfysize=3.0in
	    \epsfbox{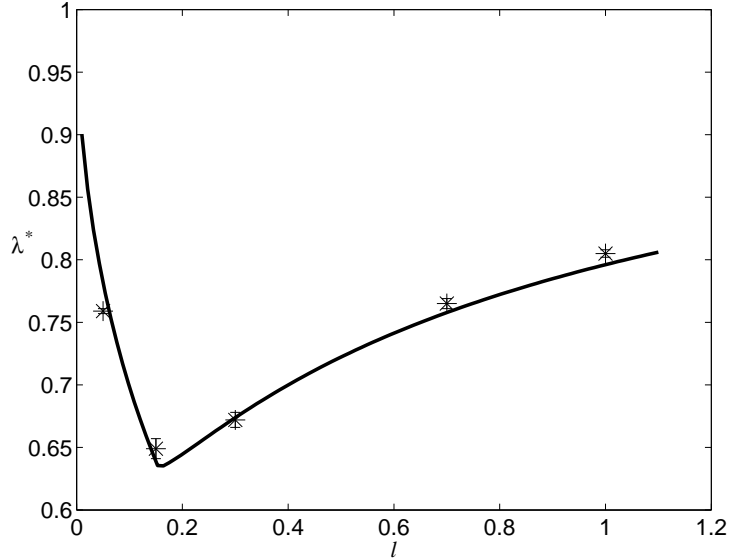}
		\caption{Variational estimate on the second eigenvalue for the one dimensional 
		Gaussian problem using the RWMH method,
		with $k=1.0$. The variational estimate is the solid line and the empirically
		determined values are marked with stars. Some of the empirical values
		seem to be less than the lower bound, but this is due to inaccuracies
		in their estimation. The optimum occurs at $l\approx 0.165$.}
	    \label{fig:optk3_oldmc}
		\end{center}
\end{figure}

Next, we use a numerical 
optimization method to maximize the bound defined by Eqn. \ref{varbound} with
respect to $a$.
The result of this analysis is shown in Fig. \ref{fig:optk3_oldmc} along
with an empirically determined convergence rate for comparison.
(To obtain the rate empirically, we run the MCMC algorithm for many iterates
on a random initial distribution and observe the pointwise differences from the distribution
of the $m^{th}$ iterate and the target distribution for large $m$. 
These differences are either fit using Hermite polynomial functions or by looking for 
the multiplicative factor which describes the geometric decay of the $m^{th}$ difference 
from one iterate to the next.)
The variational bound tightly matches the empirical obtained eigenvalue estimates 
in this case, and an optimum step size $l$ can be ascertained. Clearly
our $l=1$ initial guess for the best scaling is far from optimal. 

It is also worth comparing the optimal step scale with those obtained from different methods.
In~\citep{weakconvergence,optimalscaling}, a derivation of the optimal step size and 
acceptance rate is proposed based on minimizing the integrated autocorrelation time of an arbitrary
function of the chain's states in stationarity. By approximating the chain as an infinite
dimensional diffusion process, formulas are derived for the optimal scaling of
steps. 
For our one dimensional Gaussian target density, the proposal density's optimal variance 
is suggested to be $(1/2.38)^2 = 0.176$ which is surprisingly close to the estimate we
have obtained using the variational method from Fig.~\ref{fig:optk3_oldmc}, $l \approx 0.165$. 
However, given the infinite dimensional limit in which the former approximation is 
made, and the different convergence criterion based on autocorrelation time rather than
second eigenvalue, the agreement may be merely coincidental.

Moving to the one dimensional smart Monte Carlo, we have a Gaussian proposal 
density of the form : 
\begin{equation}
q(x,y) = \sqrt{\frac{l}{2\pi}} \exp 
\left(-\frac{1}{2} l \left( y-(x-\frac{k}{h} x) \right)^2 \right)
\label{1dqxysmart}
\end{equation}
where $1/l$ is the variance of the random part of the step and $1/h$ 
is the scale of the deterministic part. (Letting $h\to\infty$ we recover
the RWMH algorithm of Eqn. \ref{1dqxy}.) 

Taking $h=k$ corresponds to performing a Newton step at 
every iterate of the algorithm. Thus, since the log of the target 
density is purely quadratic, the current point will always be 
returned to the extremum at $0$ by the deterministic component of the
smart Monte Carlo step and the random component will give a combined 
move drawn from $q(x,y) = q(y) = \sqrt{l/(2\pi)} \exp\left(-l y^2/2\right) $, 
which has the form of an \emph{independence sampler}~\citep{casella}. 
If we then 
also choose $l = k$, we see immediately that we are generating 
moves from the target distribution from the beginning, i.e. we have convergence
in one step starting from \emph{any} initial distribution.

In real problems, however, $-\log(\pi(x))$ will not be quadratic. We 
may obtain an estimate for $l$ and $h$ by considering its quadratic
approximation or curvature but in many cases those estimates will 
have to be adjusted. If the curvature is very small (or in multidimensional
problems if the quadratic approximations are close to singular), 
the parameters will have to be increased to provide a step
size control to prevent wildly unconstrained moves (analogous to the
application of a trust region in optimization methods~\citep{trustregion}).
If the curvature is large but we believe that the target density is 
multimodal, we need to decrease the parameters to allow larger steps
to escape the local extrema. Therefore we examine in the following 
the dependence of the convergence rate as we vary both of the parameters
$l$ and $h$.

\begin{figure}[hbtp]
  \vspace{2pt}
  
\centerline{\hbox{ \hspace{0.0in}
    \epsfxsize=1.7in
    \epsffile{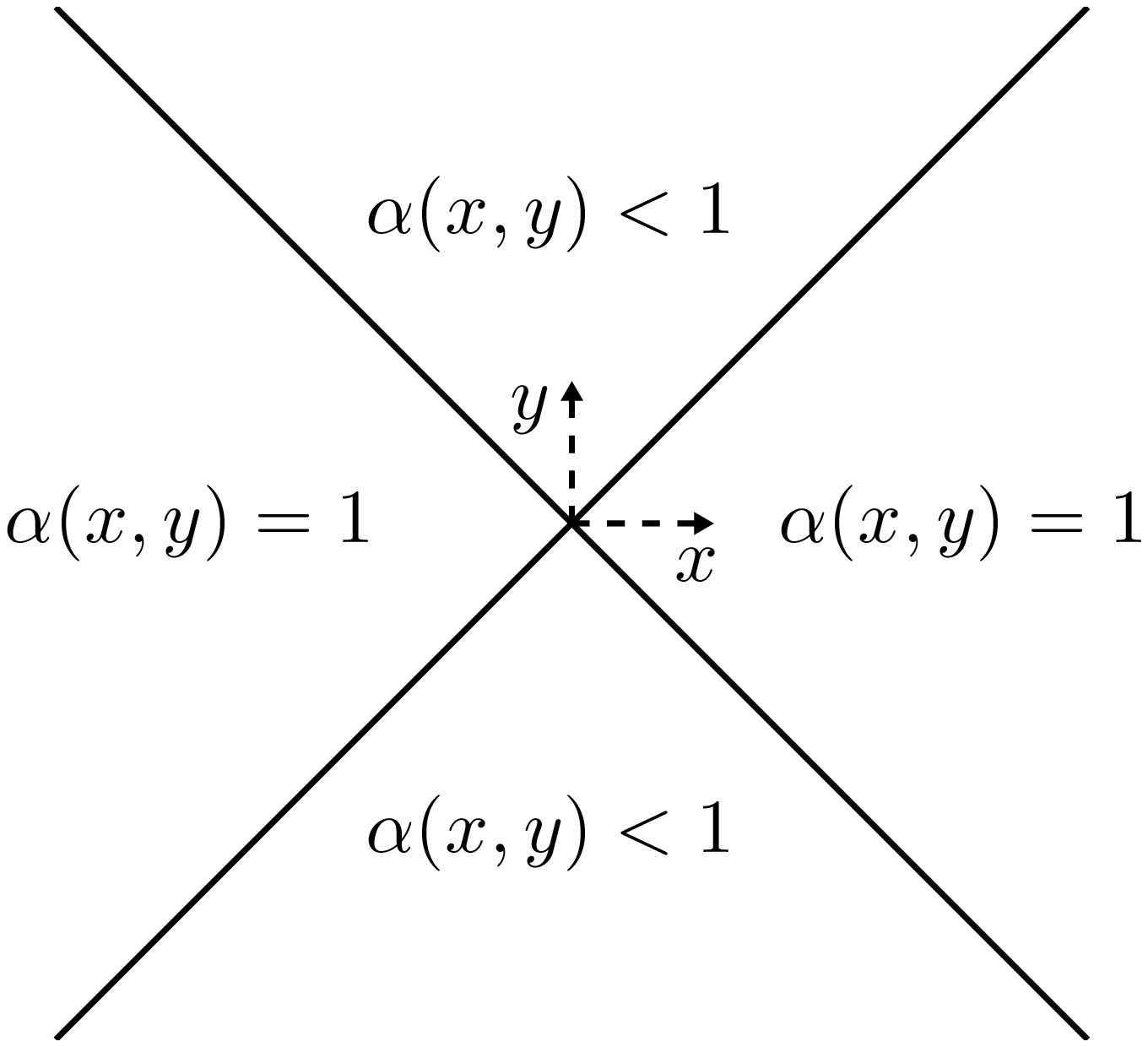}
    \hspace{0.15in}
    \epsfxsize=1.7in
    \epsffile{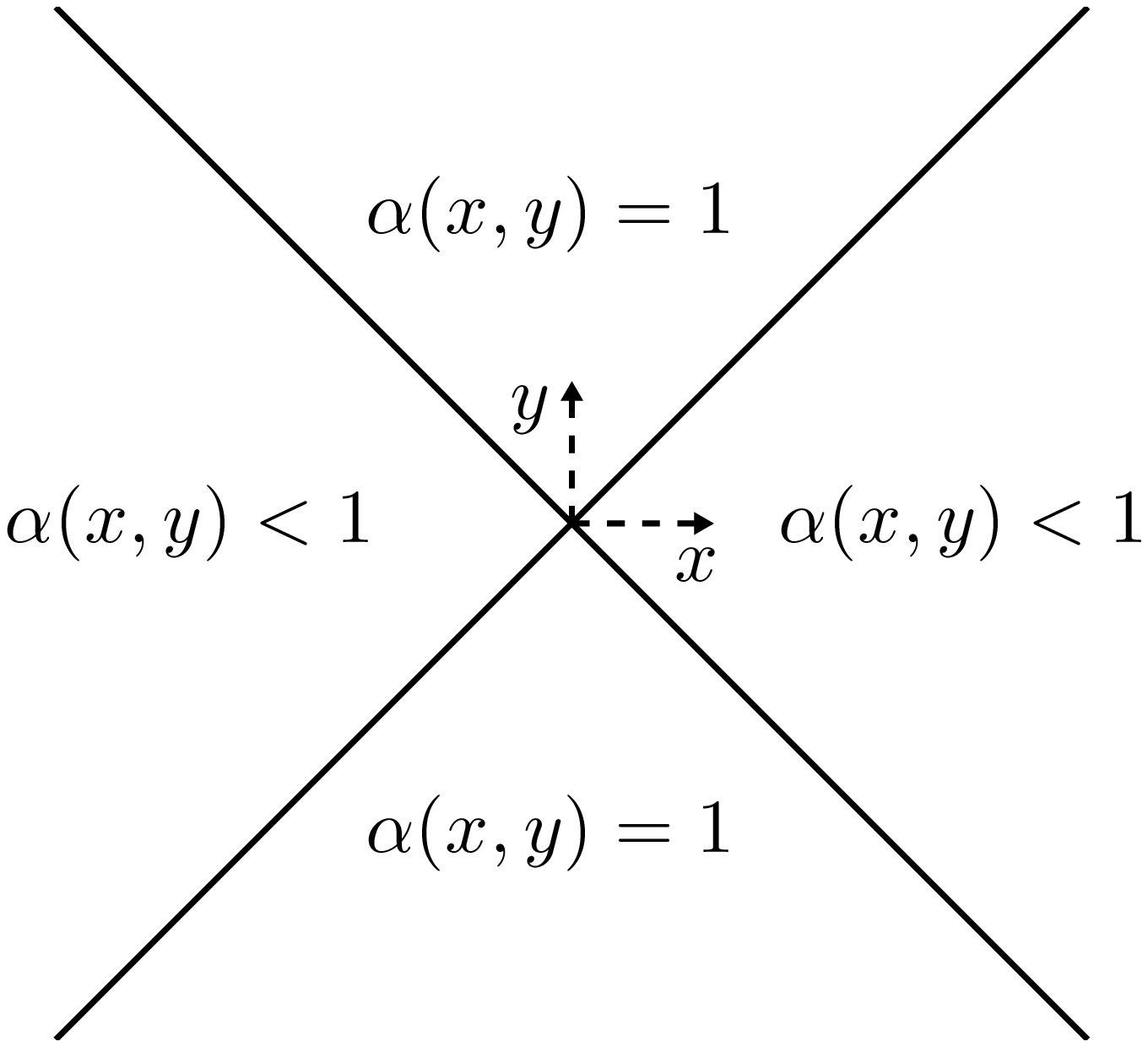}
  	\hspace{0.3in}	
	\epsfxsize=1.8in
	\epsffile{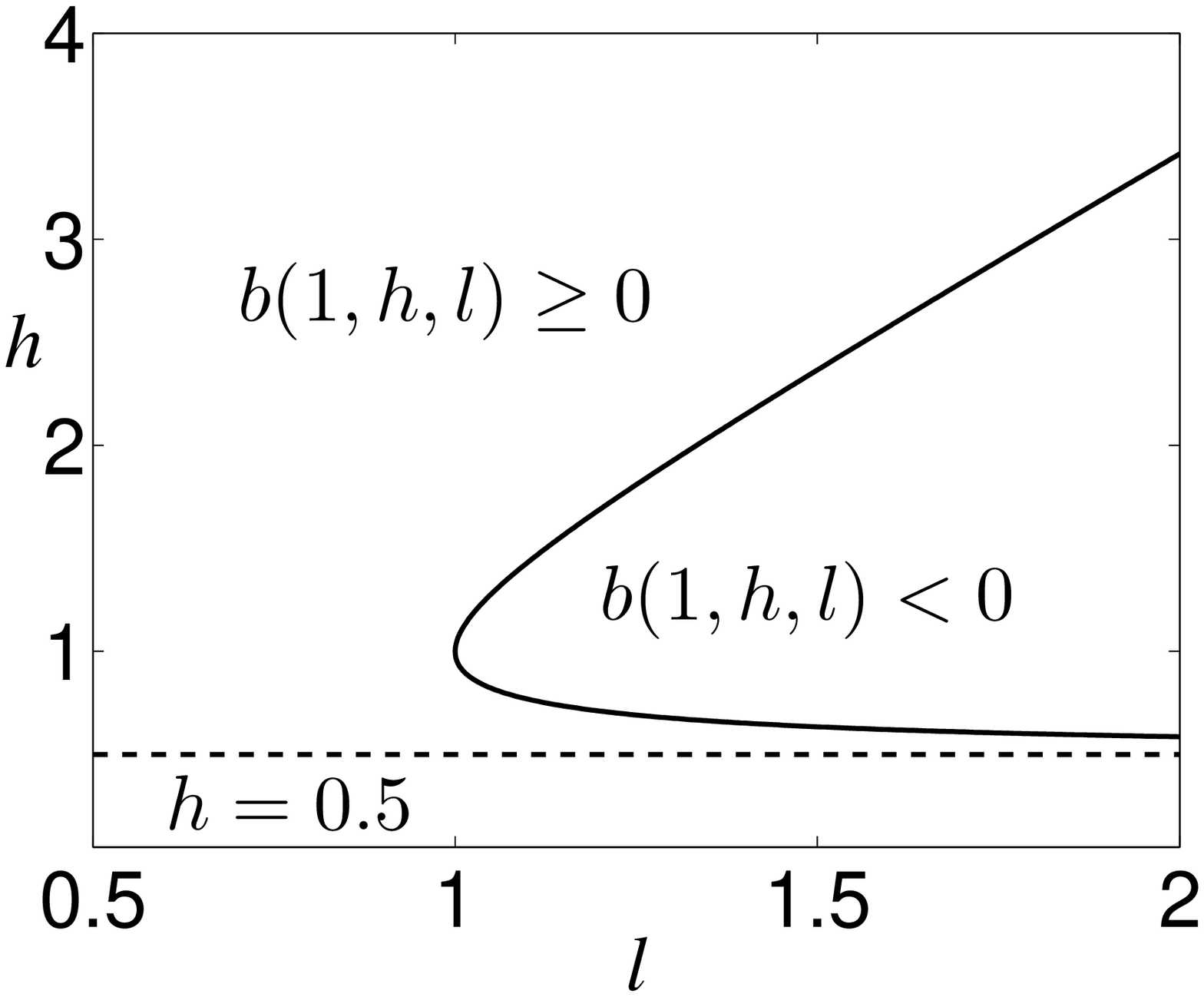}
   }
 }
\vspace{12pt}
  \hbox{\hspace{.4in} (a) \hspace{1.65in} (b) \hspace{2.0in} (c)}
  \caption{Regions in xy plane where acceptance probability $\alpha(x,y)<1$
	or $\alpha(x,y)=1$, when (a) $b(1,h,l)\ge 0$ and 
	(b) $b(1,h,l)< 0$. The
	equation for the boundary is shown in (c), 
	see Eqn. \ref{boundary} with $k=1.0$. (The RWMH algorithm will
	only have regions described by (a).) } 
  \label{fig:xy_regions}
\end{figure}

The acceptance probability Eqn. \ref{acceptanceratio} has two 
functional forms separated by a boundary in the $(x,y)$ plane  
given by 
\begin{equation}
\left( k + l \frac{k}{h} \left(-2 + \frac{k}{h}\right) \right) (y^2 - x^2) = 
b(k,h,l) (y^2 - x^2) = 0 \enspace . 
\label{boundary}
\end{equation}  
In particular, the acceptance probability is 
\begin{equation} 
\alpha(x,y) = \mbox{min}\left(
\exp\left( -\frac{1}{2} b(k,h,l) (y^2-x^2) \right),1 
\right) \enspace . 
\label{acceptancesmc}
\end{equation}
Now we have a complication over the RWMH method
because depending on the sign of the coefficient function $b(k,h,l)$ in 
Eqn. \ref{boundary}, we find 
that either $\alpha(x,y)<1$ on $|y|\ge|x|$, $\alpha(x,y)=1$ on $|y|<|x|$ or 
vice versa. This is shown in Fig. \ref{fig:xy_regions}.

\begin{figure}[hbtp]
  \vspace{2pt}
  
\centerline{\hbox{ \hspace{0.0in}
    \epsfxsize=3.0in
    \epsffile{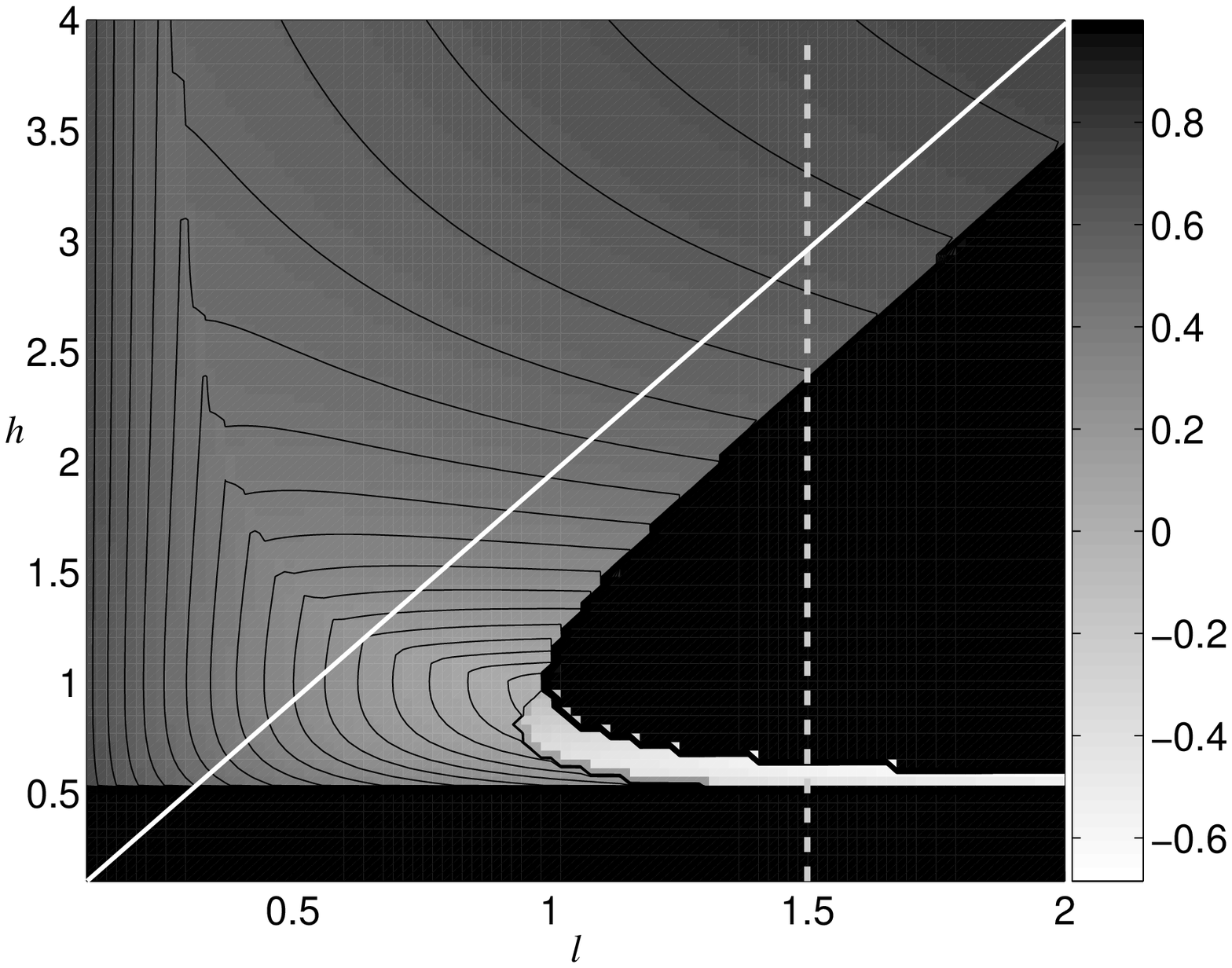}
    \hspace{0.3in}
    \epsfxsize=3.0in
    \epsffile{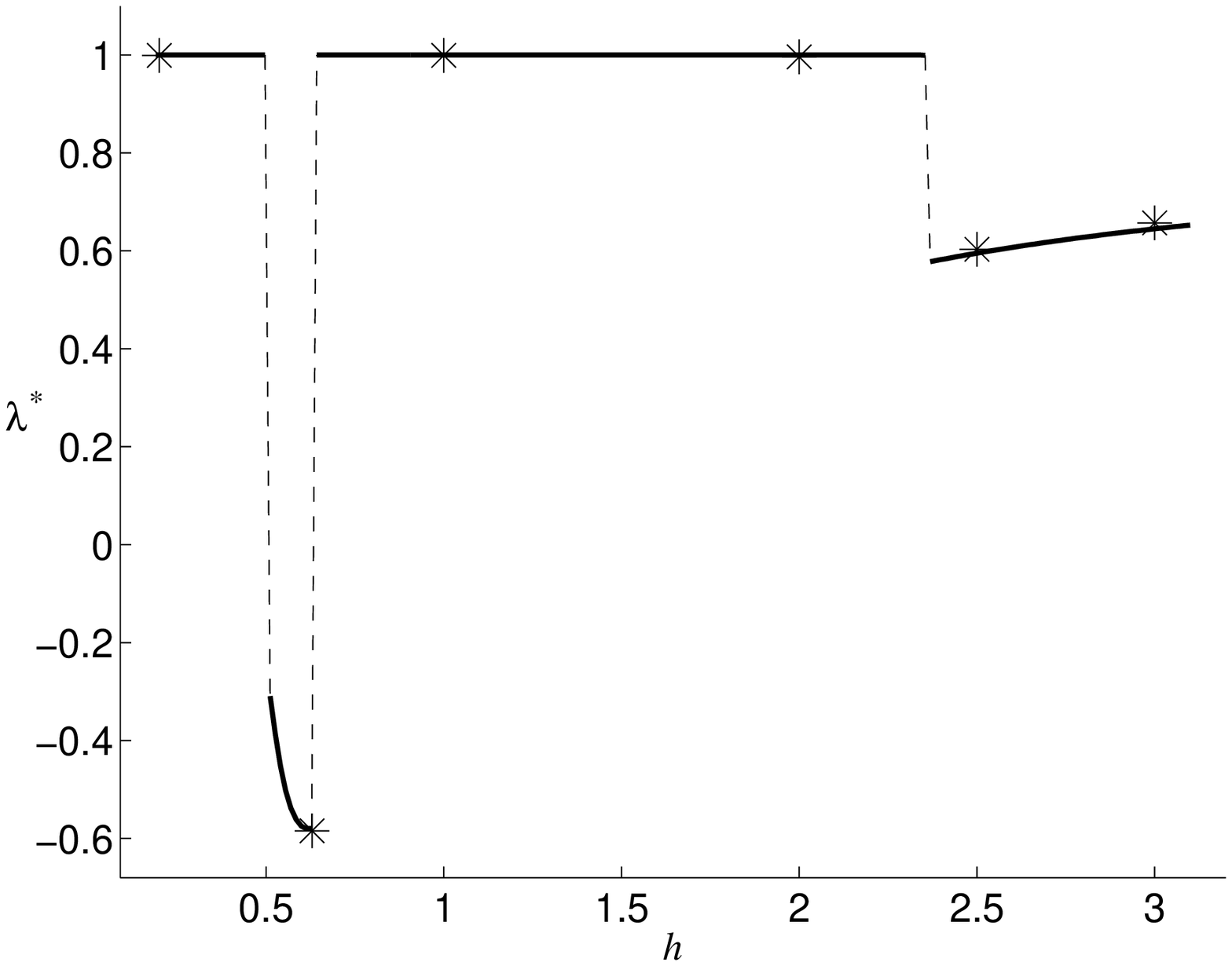}
   }
 }
\vspace{5pt}
  \hbox{\hspace{.6in} (a) \hspace{3.4in} (b)}
  
	\caption{Estimate of second eigenvalue for the symmetrized 
	smart Monte Carlo operator. (a) $k=1$ is fixed and $h$, $l$ are 
	allowed to vary. $h = 1.0, l = 1.0$ is the optimal scaling for 
	deterministic and random parts of the step. The solid diagonal line is
	the parameter restrictions that yield the MALA algorithm (see text). 
	(b) We take a slice
	through this surface at $l = 1.5$ and empirically determine the second
	eigenvalue at points along this curve (stars). The error bars are too small to be seen.
	Dashed lines are discontinuities.
	}
  \label{fig:quadevals}
\end{figure}

As before, for a given value of $h$ and $l$, we need to break up 
the double integrals of the scalar product $(v_a,\mathcal{S}v_a)$, Eqn. \ref{scalarprod}, 
into the appropriate regions.
Our choice of variational function is the same as before (since the target
density is the same) and we again can get an explicit (but complicated)
expression for Eqn. \ref{scalarprod} which we maximize with respect to $a$.
The results of this analysis are shown in
Fig. \ref{fig:quadevals} (a), where we fix $k=1.0$ and vary $h$, $l$.
We have confirmed that these lower bounds are quite accurate as shown 
in Fig. \ref{fig:quadevals} (b). 

The remarkable feature of these results is that even for this simple Gaussian 
problem, the selection of step scale parameters $h$, $l$ is critical to achieve convergence. 
As already mentioned, there is a trivial choice of optimum with $h = l = k = 1$ that 
gives one step convergence from any initial distribution (and therefore 
$\lambda^* = 0$). However, if we change parameters infinitesimally such that 
$l = 1+\epsilon, h = 1$ ($\epsilon>0$) we go through a discontinuous transition where we see 
no convergence from \emph{any} initial distribution. This can be understood
by recognizing that after one step we will have a proposal density 
(before accept/reject) $\propto \exp(-(1+\epsilon) x^2/2)$ which has a factor 
$\exp(-\epsilon x^2/2)$ less probability in its tails than the target 
density. Suppose there is an initial distribution or point mass concentrated at 
$x = \sqrt{2 M}/\sqrt{\epsilon}$, $M>>1$. The proposed step of the smart Monte Carlo algorithm, starting
at $x$, will revisit $x$ too infrequently by a factor $\exp(-M)$. Thus detailed balance will 
force the transition $x\to0$ to be accepted with a probability of only $\exp(-M)$, and thus the initial 
distribution will take an arbitrarily long time to converge to the target density.

More formally, we can compute the probability of rejection, $r(x) = 1 - \int t(x,y) dy$, when $h,l$ are as above and we find,
\[ r(x) = \Phi \left( \sqrt{\frac{1+\epsilon}{2}} x \right) - \exp \left( \frac{1}{2} b(k,h,l) x^2 \right) \frac{1+\epsilon}{1+\epsilon+b(k,h,l)} \Phi \left( \frac{1+b(k,h,l)+\epsilon}{2} x \right) \]
where $b(k,h,l) < 0$ and $\Phi$ is the cumulative normal $(0,1)$ distribution function.
We note that $\mbox{ess sup } r(x) = 1$ by continuity of $r(x)$, and then we use Proposition $5.1$ 
from~\citep{geometricconvergence} to conclude that the Markov chain is no longer geometrically 
convergent for these values of $h$ and $l$.

\begin{figure}[hbtp]
		\begin{center}
		\epsfysize=3.0in
	    \epsfbox{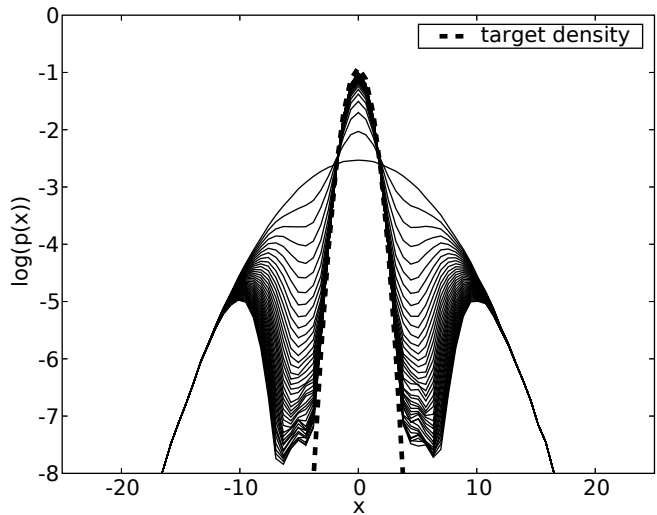}
		\caption{Forty iterates of the smart Monte Carlo algorithm (solid lines), 
		Eqn. \ref{1dqxysmart}, 
		when the initial distribution is normal with standard deviation
		five times the Gaussian target density (dashed line). Parameters are chosen to 
		be in the region of no convergence ($h = 2.0$, $l = 1.5$), see 
		Fig. \ref{fig:quadevals} (a).
		We see that the tails of the initial distribution are essentially
		unchanging after many iterates and have failed to converge to 
		the target density} 
	    \label{fig:NoConQuad}
		\end{center}
\end{figure}

In fact this is only one of the two 
disconnected regions where no convergence is observed in 
Fig. \ref{fig:quadevals}.
The largest of the two (with $h> 1/2$) is defined exactly by the 
equation $b(1,h,l)<0$ (compare Fig.~\ref{fig:xy_regions}(c) with 
Fig.~\ref{fig:quadevals} (a)). 
In this region the bound 
on the second eigenvalue approaches $1$ as the variational parameter, $a \to 0$. 
This corresponds to a perturbation on the target density of $x \sqrt{\pi(x)}$ 
for the unsymmetrized MCMC operator $\mathcal{L}$. 
In other words, we have a test distribution that has exponentially more probability
in its tails than the target density. For initial states $x$ 
arbitrarily far away from the origin, 
the acceptance probability $\alpha(x,y)$ in the region $|y|<|x|$ is
arbitrarily small. To see this, note that Eqn. \ref{acceptancesmc} is an
exponentially decaying function of $y^2-x^2$ in this region, and given the form
of the proposal density Eqn. \ref{1dqxysmart}, we see that the expected value
of $y^2-x^2$ is arbitrarily large and negative.
Thus states far out will never be ``allowed back'' and the fat tails of $\sqrt{\pi(x)}$
will never shrink back down those of $\pi(x)$. Furthermore, moves $x\to y$ where 
$|y|\ge|x|$ are always accepted (because $\alpha(x,y)=1$ on $|y|>|x|$) which 
simultaneously prevents
convergence. The situation is analogous to that described for $l=1+\epsilon$ and $h = k = 1$, 
except now there is a cutoff both on the deterministic step and the random step. 
A typical example of this is shown in Fig. \ref{fig:NoConQuad}. 
Once we cross to the $b(1,h,l)\ge0$ region, moves $x\to y$ where $|y|<|x|$
are always accepted by Eqn. \ref{acceptancesmc} (Fig. \ref{fig:xy_regions} (a)). 
Therefore excess probability in the tails is allowed to flow back into the central part of the distribution 
and the convergence is not blocked.

In the second region where no convergence is observed, ($h< 1/2$ in Fig. 
\ref{fig:quadevals}), we have a situation where the deterministic step alone 
(taking $l \to \infty$) leads to the proposed moves being generated by an  
unstable mapping, from the $(n-1)^{th}$ to
$n^{th}$ iterate: $x^{(n)} = x^{(n-1)} - \beta x^{(n-1)} $ where $\beta>2$. 
The trial variational function for this situation also maximizes the bound 
as $a\to0$, again implying that the tails are not decaying to the stationary
distribution. 
The reason is that, even when $l < \infty$, we have a situation
in which the \emph{expected} or mean position of a state $x$ after one step
is $y$ where $|y|\ge|x|$. Thus excessive probability in the tails cannot be shifted
inward to match the target density. 

The lack of convergence in this region was already noted for the Metropolis Adjusted
Langevin Algorithm (MALA), a special case of the SMC algorithm where $h = 2l$. As shown in~\citep{mala},
if $\pi$ is bounded, a sufficient condition for MALA to fail to be geometrically ergodic is 
\[ \liminf_{|x|\to \infty} \frac{|\nabla \log \pi(x)|}{|x|} > \frac{4}{s} \] where $s$ is the single stepsize
control parameter for that algorithm. The equivalence to the SMC method is established by setting $l = 1/s$.
Thus, for the Gaussian target density $\pi$, the condition is $l < 1/4$. Referring to the 
solid white line in Fig.~\ref{fig:quadevals}(a), 
the non-convergent parameter regime for MALA lies along the line segment $h=2l$ with $l < 1/4$ which matches exactly with the boundary we have determined using the variational method.

The $h=1/2$ ``trough'' is a special case where we have
oscillatory behavior. That is, the second eigenvalue is negative but greater 
than $-1$ and in fact convergence does occur. Interestingly setting $h = k/2$ means
that $b(k,h,l) = k$ and the acceptance probability of Eqn. \ref{acceptancesmc}
looks again like that of the RWMH algorithm, but the convergence is actually
faster. In a sense, given that the deterministic part of the step moves $x\to-x$ and 
the target distribution is symmetric, the oscillatory behavior allows the chain
to sample the distribution twice as fast.

\subsection{Quartic target density}
In scientific or statistical applications where MCMC is used, the log of the target 
density will ordinarily have higher order terms beyond the quadratic order we studied 
in the previous section. For example, in a Bayesian inference problem the posterior distribution  
will rarely have a simple Gaussian form. 
Both finding the maximum \textit{a posteriori} parameter estimates and sampling from the posterior are
made more difficult in the presence of these higher order terms.

Therefore, we wish to extend the previous example by studying a target distribution of the form 
$\pi(x) = (2^{(3/4)} k^{(1/4)} /\Gamma(1/4) )
\exp\left(- k x^4 /2 \right)$.
Here, the log of the target density is quartic and the proposal density (Gaussian) no 
longer has the same form as the target density. We would 
like to understand the performance of the Monte Carlo algorithms in this circumstance.
(The test distribution is taken to be $\propto x \exp\left(- a x^4 /2 
\right) $, i.e. in the orthogonal space to the stationary distribution).

\begin{figure}[hbtp]
		\begin{center}
		\epsfysize=3.0in
	    \epsfbox{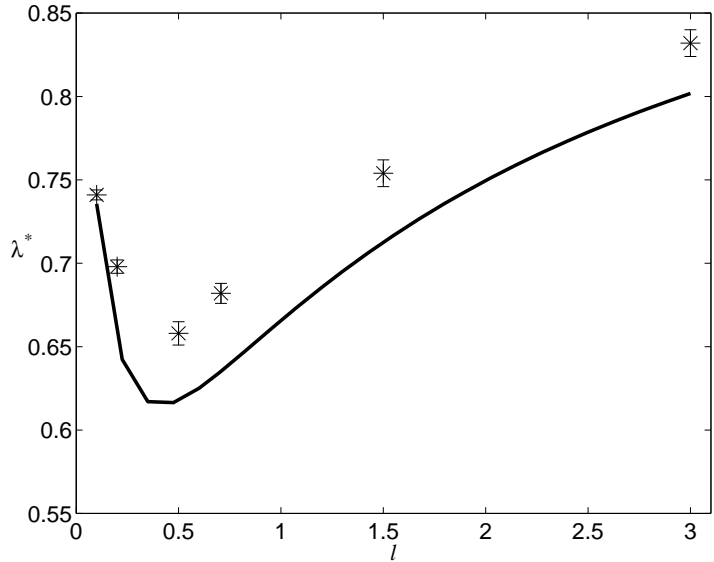}
		\caption{Second eigenvalue estimate from the variational method (solid line)
		and empirical estimates (stars), for the quartic target density ($k=1$) using 
		the RWMH method, Eqn. \ref{1dqxy}.  
		The numerical values for $\lambda^*$ are now estimated by taking the ratio
		of the discrepancy from the target density in subsequent iterates 
		and finding a single multiplicative factor which describes the decay. 
		This is done rather than using functional forms analogous to Hermite 
		polynomials to fit the decay, because it appears that there may be more 
		significant contributions from higher order terms. This also explains why 
		the lower bound shown differs more than in Fig. \ref{fig:optk3_oldmc}
		and Fig. \ref{fig:quadevals} (b). The data point shown at $l = 1/\sqrt{2}\approx .71$ (see text)
		does not appear to be optimal. }
	    \label{fig:oldmcConQuart}
		\end{center}
\end{figure}

The goal is to estimate the optimal value of $l$, as before. We can argue 
approximately that the step scale should be such that $ k x^4 /2 \approx 1$ for a typical move $x$,
i.e. the change in energy is about $1$ and the acceptance probability is therefore
$\exp(-1)$. This gives a typical value for $x^2 = \sqrt{2}/\sqrt{k}$. 
Since the proposal density is Gaussian with variance $1/l$, we therefore 
would naively predict $l=\sqrt{k}/\sqrt{2}$. 
Applying the variational method, we were unable to find a closed form 
solution to Eqn. \ref{scalarprod} so we had to resort to numerical integrals in 
determining the bound in Eqn. \ref{varbound}. The results are shown in 
Fig. \ref{fig:oldmcConQuart} for the RWMH method;  
it suggests an optimal choice for the step size parameter, $l$, which 
is an improvement over our initial guess of $1/\sqrt{2} \approx .71$ (when $k = 1$). 
Using the formulas for the optimal step scale from~\citep{optimalscaling} coincidentally yields
about $.71$, also a little off from our variational estimate.

Turning to the smart Monte Carlo algorithm, if we wish to make the deterministic part 
of the proposed move a Newton step using the Hessian of $-\log(\pi(x))$ at $x=0$ we 
are left with a singular Hessian and an infinite deterministic step, reinforcing the need 
for the step length control parameter, $h$.

\begin{figure}[hbtp]
		\begin{center}
		\epsfysize=3.0in
	    \epsfbox{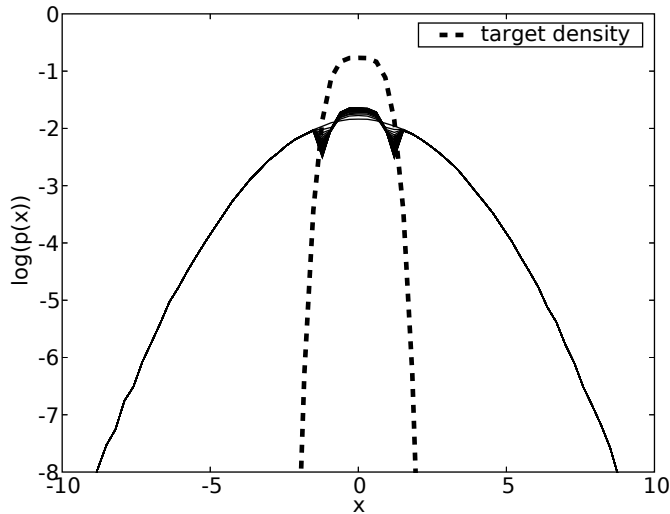}
		\caption{Forty iterates of the smart Monte Carlo algorithm (solid lines),  
		Eqn. \ref{1dqxysmart}, when the target density is quartic (dashed line). 
		The initial distribution of points is normal with standard deviation
		about five times that of target density (dashed line). Parameters are 
		arbitrarily chosen as ($h = 1.0$, $l = 1.0$), and
		we see that the tails of the initial distribution are unchanged for 
		every iterate of the algorithm.
		Other parameter sets tested lead to the same behavior.
		} 
	    \label{fig:NoConQuart}
		\end{center}
\end{figure}

Surprisingly, we find that, \emph{independent} of the value of $h$ and $l$,
($k$ fixed at $1$), the scalar product $(v_a,\mathcal{S}v_a) \to 1$ as $a \to 0$. Thus 
there are no choices of scaling parameters which will lead to convergence. 
This is borne out by numerical simulation, see Fig. \ref{fig:NoConQuart} for 
the changes in an initial density under many iterates of the algorithm with an
arbitrary choice for $s$, $h$.

The failure of the smart Monte Carlo method for the quartic problem is clearly 
due to non-convergence of the tails of the distribution, and can be seen 
by analyzing the integrals defining the operator, Eqn. \ref{scalarprod}, and noting
that they all tend to zero as the variational parameter tends to zero, independent of the choice
for $k$, $h$ and $l$. 

As a partial check on this result, we again apply the condition derived 
in~\citep{mala} for the MALA algorithm, which states that geometric
convergence is not possible when 
\linebreak[4]$\liminf_{|x|\to \infty} |\nabla \log \pi(x)|/|x| > 4/s$ 
where $s$ is the step scale parameter. Now, for the quartic density $\pi$, the quantity on the
left of the inequality $\to \infty$, so \emph{no} value of $s$ can give geometric convergence. 
MALA is a special case of the SMC algorithm, but we have shown here that the latter also has
convergence problems indicated by $\lambda^* \to 1$, for \emph{all} values of its scaling
parameters, $h$ and $l$.

The Gaussian and quartic problems are representative examples of target densities
on which we have tested the smart Monte Carlo method. As we have seen there are 
severe convergence problems on these distributions. We would expect that for real 
applications, where the log of the target density would contain components of these and 
higher order nonlinearities, similar convergence difficulties for the smart Monte Carlo 
method would occur. It may well be that in applications where the method is extensively used 
({e.g.}~\citep{CHARMM,adsorption,transportcoeffs}) the convergence criteria are less precise
than ours. For example, it may be acceptable to merely monitor the variance of some function 
of the state space variables and conclude that convergence has been achieved when it ceases to
change appreciably, or as in~\citep{optimalscaling}, define efficiency by the integrated 
autocorrelation time.

\section{Multidimensional target densities}
For multidimensional problems, it is quite common to find a large range of 
scales associated with the target density~\citep{kevin1,Ryan,Josh}. That is, the curvature
of the probability density along some directions in the parameter space is much larger than
in other directions. Clearly, if an MCMC method is not designed to take these different 
scales into account through importance sampling, the algorithm will perform very poorly. If the
curvature is very high in a particular direction and we try to take a moderately sized step, 
it will almost certainly be rejected but if we take small steps in directions that are 
essentially flat the MCMC algorithm will be very slow to equilibrate. We would like
to show explicitly here what happens to the convergence rate when the scale of the 
problem has been underestimated or overestimated.

The variational calculations for the one dimensional examples of the 
previous section either yielded explicit formulas or gave integrals that were relatively 
fast to compute numerically. However as we go to multiple dimensions neither of these 
features are present, in general. 
Typically the integrals describing $(v_a,\mathcal{S}v_a)$ will \emph{not}
factor into one dimensional integrals. For Gaussian target densities the full
space is broken into regions analogous to those in Fig. \ref{fig:xy_regions}, 
described by an equation like $y^t A y \ge x^t A x$ where $A$ is a 
symmetric $n$ by $n$ matrix which is not necessarily
positive definite. For the RWMH algorithm applied to a multivariate Gaussian
target density with inverse covariance matrix $K$, we have $A = K$, and 
therefore all the dimensions are coupled through the energy change, $\Delta E = y^t K y - x^t K x$.
We would still like to be able to get a lower bound on $\lambda^*$, and to this end
note that \emph{any} test function orthogonal to the target density will 
work in Eqn. \ref{varbound}; the bound does not explicitly 
require a variational parameter, however without it the estimate will be less accurate.
It is still necessary to make choices for the test functions that are both tractable
in computing $(v,\mathcal{S}v)$  and are ``difficult'' functions for the given algorithm
to converge from, i.e. have a significant component along the true second eigenfunction.

As an example, take the multivariate Gaussian distribution of the form
\begin{equation}
\pi(x) = \frac{\sqrt{|K|} }{(2 \pi)^\frac{n}{2}} 
\exp \left(-\frac{1}{2} x^t K x \right)
\label{multivar_dens}
\end{equation}
with $x = (x_1,...,x_n)$,  
and consider using the MH algorithm with importance sampling, i.e.
\[ q(x,y) = \frac{\sqrt{|L|} }{(2 \pi)^\frac{n}{2}}
\exp\left(-\frac{1}{2} (y-x)^t L (y-x) \right) \]
where again $L$ is the inverse covariance matrix/step size control term and to simplify 
we assume that both $K$ and $L$ are diagonal.
Without any analysis we might guess that the optimum choice for $L$ is $K$. 

First we construct a test function that will provide a useful bound when the proposed steps 
are too large for the natural scale of the problem. For simplicity, consider putting a 
delta function distribution at the origin. If we take large steps the 
acceptance probability should be low and there will be a large overlap between the initial
state and the final state. In the limit that the proposed steps have infinite length, the
initial state will not be changed at all and the bound on the second eigenvalue in absolute
value will approach one. To do this more carefully we define a test function which is a Gaussian 
whose variance will ultimately be taken to zero to represent the delta function. However, we also need to add 
another term to ensure the test function is orthogonal to the target density, in order to apply
the variational bound.
Therefore, for the unsymmetric operator we write the test function as : 
$u_\sigma(x) = -A \pi(x) + B w_\sigma(x) $ where $w_\sigma(x)$ is the probability density
for a multivariate Gaussian with covariance matrix $\sigma^2 I$ and $A$ and $B$ are 
constants. 
For the symmetrized operator the trial function is transformed to  
$v_\sigma(x) = -A \sqrt{\pi(x)} + B w_\sigma(x)/\sqrt{\pi(x)} $. $A$ and $B$ are
constrained to satisfy the orthogonality relation 
$(v_\sigma,\pi) =  0$ and a normalization $(v_\sigma,v_\sigma) = 1$. These lead to the 
conditions
\begin{equation*}
A = B 
\quad \mbox{ and } \quad 
B^2 \int \left( \frac{w_\sigma(x)}{\sqrt{\pi(x)}} \right)^2 \,dx = 1 + B^2 \enspace . 
\end{equation*}
Then it can be seen that 
\[ (v_\sigma,\mathcal{S} v_\sigma) = -B^2 + 
B^2 \left(\mathcal{S} \frac{w_\sigma(x)}{\sqrt{\pi(x)}} , \frac{w_\sigma(x)}{\sqrt{\pi(x)}} \right) \]
where we have used the orthogonality condition, the fact that $w_\sigma(x)$ integrates to $1$
and that $\mathcal{S}$ is self-adjoint. Writing out the operator $\mathcal{S}$ explicitly we get
\begin{eqnarray*} 
\left(\mathcal{S} \frac{w_\sigma(x)}{\sqrt{\pi(x)}} , \frac{w_\sigma(x)}{\sqrt{\pi(x)}} \right) & = & 
\int \int \frac{w_\sigma(x)}{\sqrt{\pi(x)}} s(x,y) \frac{w_\sigma(y)}{\sqrt{\pi(y)}} \, dx dy -
\int \int t(x,y) \left( \frac{w_\sigma(x)}{\sqrt{\pi(x)}} \right)^2 \, dx dy \\
& & + \int \left( \frac{w_\sigma(x)}{\sqrt{\pi(x)}} \right)^2 \, dx \enspace .
\end{eqnarray*} 
The last term on the right hand side is $(1+B^2)/B^2$, making use of the normalization condition, 
so we are left with
\begin{equation*}
(v_\sigma, \mathcal{S} v_\sigma) = 
B^2 \int \int \frac{w_\sigma(x)}{\sqrt{\pi(x)}} s(x,y) \frac{w_\sigma(y)}{\sqrt{\pi(y)}} \, dx dy - 
B^2 \int \int t(x,y) \left( \frac{w_\sigma(x)}{\sqrt{\pi(x)}} \right)^2 \, dx dy  + 1 \enspace . 
\end{equation*}
Since we are ultimately taking a limit as $\sigma \to 0$ ($w_\sigma \to $ a delta
function) we can make approximations to these
integrals as follows :
\begin{equation*}
\int \int \frac{w_\sigma(x)}{\sqrt{\pi(x)}} s(x,y) 
\frac{w_\sigma(y)}{\sqrt{\pi(y)}} \, dx dy 
\approx s(0,0) \int \int 
\frac{w_\sigma(x)}{\sqrt{\pi(x)}} \frac{w_\sigma(y)}{\sqrt{\pi(y)}} \, dx dy 
\end{equation*}
and 
\begin{equation*}
\int \int t(x,y) \left( \frac{w_\sigma(x)}{\sqrt{\pi(x)}} \right)^2 \, dx dy \approx
\int t(0,y) \, dy \int \left( \frac{w_\sigma(x)}{\sqrt{\pi(x)}} \right)^2 \, dx \enspace . 
\end{equation*}
Finally by taking $\sigma \to 0$ we have the expression 
\begin{equation*} 
(\mathcal{S}v_0, v_0) = 1 - \int t(0,y) \, dy \enspace .
\end{equation*}
As already mentioned, for the multidimensional problem we expect different functional forms
for the kernels $s(x,y)$ and $t(x,y)$ depending on the initial and final state $(x,y)$ and
this is what makes decoupling the integrals difficult. However for this choice of test
function the equation for the boundary (with $x = 0$) is given by
$y^t K y = 0$ and since $K$ is positive semidefinite
we always stay on one side of the boundary (the energy never decreases from 
the initial distribution placed at $x=0$). Then 
\begin{eqnarray}
(\mathcal{S}v_0, v_0) & = & 1 - \frac{\sqrt{|L|} }{(2 \pi)^\frac{n}{2}} \int \exp \left(
-\frac{1}{2} y^t (K + L) y 
\right) \, dy \\
			& = & 1 - \prod_{i=1}^{n} \sqrt{\frac{l_i}{l_i + k_i}} \enspace . 
\label{bigstepbound}
\end{eqnarray}
where $l_i$ and $k_i$ are the diagonal elements of the diagonal matrices $L$ and 
$K$, respectively. With no importance sampling we would have $L = k I$ where
$k$ would be chosen to make sufficiently large steps to enable it to sample $\pi(x)$. 
A rough argument as follows can give some insight into the form of Eqn. \ref{bigstepbound} :
$1/\sqrt{l_i}$ is a measure of the scale in the $i^{th}$ coordinate direction of the proposal density, 
$1/\sqrt{k_i}$ is the scale in the $i^{th}$ coordinate direction of the target density. Suppose that
$l_i \ll k_i$ for each $i$, that is the scales of the proposal density are too large in all 
directions. Then the ratio of the mean volume of moves generated by $q(0,y)$ to the volume
occupied by $\pi(y)$ is exactly $\prod_{i=1}^{n} \sqrt{l_i}/\sqrt{k_i}$. 
Intuitively, this ratio is proportional to the acceptance probability, and in the 
regime $l_i \ll k_i$ the acceptance probability determines the convergence 
properties.

We want to use Eqn. \ref{bigstepbound} to show how choosing step sizes \emph{too large}
even in one direction will result in a very inefficient algorithm. 
Suppose that for all but one of the directions we make $l_i=k_i$, $i = 1,\ldots,n-1$ 
which would be roughly the correct scaling in those directions.
Then the bound on the 
second eigenvalue is 
\begin{equation}
(\mathcal{S}v_0, v_0) =  1 - \sqrt{\left( \frac{1}{2} \right)^{n-1} } \sqrt{ \frac{1}{1+k_n/l_n} } \enspace .
\label{largestepbad}
\end{equation}
From this we can see that as we go to larger and larger step sizes relative to the scale in the
last direction ($k_n/l_n \to \infty$), the bound on $\lambda^*$ increases to $1$. 
Conversely we can argue that if one of the directions of the target density has a scale that is 
considerably smaller than the step scales being used in the proposal density, 
we will get very
few acceptances and the convergence rate will be close to $0$. Hence we see explicitly the 
need for importance sampling to accelerate convergence.

\begin{figure}[hbtp]
		\begin{center}
		\epsfysize=2.0in
	    \epsfbox{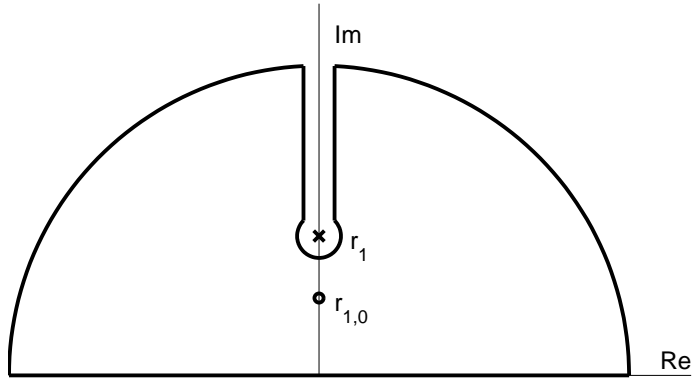}
		\caption{Contour used to evaluate Eqn. \ref{before_contour_integral} when $\Delta E < 0$.
		$r_1$ is a branch point and $r_{1,0}$ is a pole of order $(n-1)/2$. The contour is the
		same for $\Delta E <0 $ except restricted to the negative imaginary plane.}
	    \label{fig:contour}
		\end{center}
\end{figure}

\begin{figure}[hbtp]
		\begin{center}
		\epsfysize=3.0in
	    \epsfbox{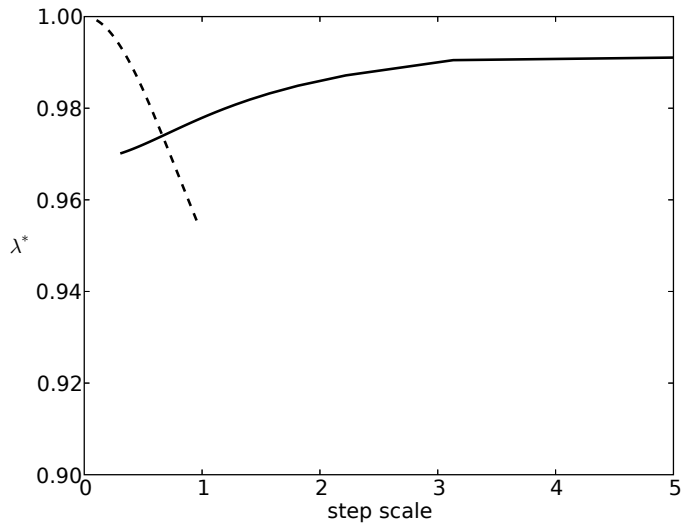}
		\caption{Lower bound on second eigenvalue for the multivariate Gaussian problem,
		Eqn. \ref{multivar_dens}, with $n=11$. Step scale =  $\sqrt{1/l_n}$. 
		$k_n=1$ sets the scale of the target density in the last direction.
		The test function is chosen as the negative of the target density 
		perturbed by a delta function (solid line) 
		or as the target density itself in all directions
		but the last (dashed line). The estimate for the lower bound is a maximum
		of the two curves.}
	    \label{fig:multi_varN11}
		\end{center}
\end{figure}

We would also like to address what happens in the other limit
as the step size becomes excessively small compared to the natural scale of the problem. (In fact
Eqn. \ref{bigstepbound} gives a lower bound of zero in that case which is not surprising as it
is based essentially on the term in the operator equation which gives the probability of \emph{staying}
at the current state. If we take infinitesimally small steps, the acceptance probability will be one and
we will never stay at the current state). When the step scales are infinitesimally small
we expect intuitively that the bound on the second eigenvalue will also approach one; 
even though the acceptance ratio is close to one, very small steps will never be able to 
``explore'' the target distribution sufficiently. 
To compute this limit, we propose a test function which has components of the target density 
in all directions except the last, where it has an antisymmetric form to make sure it 
is orthogonal to the target density. 
With respect to the symmetrized operator $\mathcal{S}$ this means 
\begin{equation}
v(x) \propto x_n \prod_{i=1}^{n} \sqrt{\pi_i(x_i)} \enspace .
\label{mulitvar_testfun}
\end{equation} 
Here $\sqrt{\pi_i(x_i)}$ is 
the one dimensional Gaussian density which is the $i^{th}$ factor in a 
diagonalized multivariate Gaussian density. (Recall that since $\pi(x)$ is an eigenfunction of $\mathcal{L}$, then $\sqrt{\pi(x)}$ is an eigenfunction of $\mathcal{S}$.)
We still have the problem of decoupling 
the $n$-dimensional multivariate problem 
into $n$ one dimensional problems. To manage this we use a device to re-express the operator equation
, Eqn. \ref{scalarprod}, explicitly in terms of the change $\frac{1}{2} (y^t K y - x^t K x)$. 
(i.e. $-\log{\frac{\pi(y)}{\pi(x)}}$), which we denote by $\Delta E$.
That is 
\begin{eqnarray*}
& (v,\mathcal{S}v) & \, = \int \int v(x) s(x,y) v(y) \, dx dy -
\int \int t(x,y) \left( v(x) \right)^2 \, dx dy + 1 \\
		& = & \int \int x_n \pi(x) q(x,y) 
		\left( \int \mbox{min}(e^{-\Delta E},1) 
		\, \delta \left( \Delta E - \frac{1}{2} \sum_{i=1}^n k_i (y_i^2 - x_i^2) \right) 
		\, d\Delta E \right) \, dx dy
		- \\ 
		&   & \int \int x_n^2 \pi(x) q(x,y) 
		\left( \int \mbox{min}(e^{-\Delta E},1) 
		\, \delta \left( \Delta E - \frac{1}{2} \sum_{i=1}^n k_i (y_i^2 - x_i^2) \right) 
		\, d\Delta E \right) \, dx dy
\end{eqnarray*}
Then we use the integral representation of the delta function 
$\delta(x) = \frac{1}{2\pi} \int \exp(-{\rm i} w x) \, dw $, factor 
$q(x,y)=\prod_{i=1}^{n} q_i(x_i,y_i)$, and rearrange the order of integration to give :
\begin{equation}
(v,\mathcal{S}v) = \frac{1}{2\pi} \int \mbox{min}(\exp(-\Delta E),1) \left(
	\int A(w) \exp(-{\rm i} w \Delta E ) \, dw \right) 
	\, d\Delta E
\label{scalarprod_with_w}
\end{equation}
where $A(w)$ contains the integration over the now decoupled $(x,y)$ coordinates :
\begin{eqnarray}
A(w) & = & 
\left( \prod_{i=1}^{n-1} \int \int \pi_i(x_i) q_i(x_i,y_i) 
\exp \left(
\frac{1}{2} {\rm i} w k_i (y_i^2 - x_i^2) \right) \, dx_i \, dy_i \right) \times \\
	&   &
\int \int (x_n y_n - x_n^2) \pi_n(x_n) q_n(x_n,y_n) 
\exp \left( 
\frac{1}{2} {\rm i} w k_i (y_n^2 - x_n^2) \right) \, dx_n \, dy_n \\ 
	& = & \prod_{i=1}^{n-1} \frac{1}{(1 + \frac{k_i}{l_i} w (-{\rm i} + w) )^{\frac{1}{2}} } \, 
	 \frac{ {\rm i} \frac{k_n}{l_n} w}{(1 + \frac{k_n}{l_n} w (-{\rm i} + w) )^{\frac{3}{2} } } 
\label{before_contour_integral}
\end{eqnarray}
Note that the complex integral with respect to $dw$ has a branch point at the roots of 
$(1 + \frac{k_n}{l_n} w (-{\rm i} + w) )^{\frac{3}{2}}$ which lie on the imaginary 
axis at $r_1$ and $r_2$. 
It simplifies the analysis to consider the situation $k_i = l_i$ for $i=1,\ldots,(n-1)$ and assume 
that $n-1$ is even. This way, the roots of $(1 + w (-{\rm i} + w) )^{\frac{n-1}{2}}$, 
$r_{1,0}$ and $r_{2,0}$, are $(n-1)/2$ order poles and not branch points, also on the imaginary axis. 
If we now also assume that $k_n < s_n$, then we can take a contour as shown in 
Fig. \ref{fig:contour} when $\Delta E < 0$ and a similar one in the lower imaginary plane
when $\Delta E > 0$.
Thus Eqn. \ref{before_contour_integral} is reduced to a residue term and 
a real integral which needs to be evaluated numerically.  
The result is plotted for $n=11$ in Fig. \ref{fig:multi_varN11} along with the bound that came from
Eqn. \ref{largestepbad}. Thus we see the trade off between taking large steps that potentially
can explore the space quickly but have a higher chance of being rejected and taking small
steps which will have a high acceptance probability but will be unable to sample the 
space quickly. 
As we saw when doing the full variational calculation for the one dimensional 
problems, the best step scale to use is not what we may have guessed;
the natural choice $l_n = k_n = 1$ here does not appear to minimize the 
second eigenvalue. 
We believe this kind of ``approximate'' variational approach may be a useful way to deal with 
problems which are difficult to analyze otherwise. 

\section{Conclusion}
By applying a variational method, it is possible to obtain an
accurate (lower bound) estimate for the second eigenvalue of an MCMC operator and
thus bound the asymptotic convergence rate of the chain to the target distribution. 
Given such an estimate we can optimally tune the parameters in the proposal distribution 
to improve the performance of the algorithm. The procedure has a role to play between 
the various numerical algorithms that perform convergence diagnostics before the full 
simulations are run, to allow the user to manually tune parameters, 
and the adaptive schemes~\citep{adaptiveschemes,atchade:0305} that require no preliminary 
exploration.
The simulations we performed to confirm our variational bounds in the case of a one dimensional
target density and varying one step scale parameter, Fig.~\ref{fig:optk3_oldmc} and Fig.~\ref{fig:quadevals}~(b), 
would be infeasible to do as we move to higher dimensions and as we vary additional algorithm parameters. 
It is in those situations that the variational method can more efficiently identify regions of optimality.

In addition, the variational method allows us to discover weaknesses in variants of 
the Random Walk Metropolis-Hastings algorithm which on the surface appear to be 
reasonable prescriptions for sampling the target density. This is most dramatically 
seen in the smart Monte Carlo method discussed above which apparently has serious flaws 
for even the simplest of one dimensional target densities. Although the smart MC method 
has been widely used in molecular dynamics 
applications~\citep{CHARMM,adsorption,transportcoeffs} the
scales are often chosen by physical considerations (for example, to not exceed significantly the 
step sizes needed to accurately describe the dynamical evolution of the system) and furthermore, 
the diagnostics of convergence are not as rigorous as ours; typically a physical quantity
is monitored till it appears to reach an equilibrium value, the rare events 
which correspond to the tails of the target distribution are possibly of lesser
importance in those studies. Therefore the convergence problems we have discussed 
here specifically in relation to the smart Monte Carlo method, to our knowledge, have 
not been previously examined.
Presumably the convergence problems can be corrected by a more careful discretization of the 
underlying diffusion equations, as was shown for the related Langevin-type methods~\citep{langevin}.

It would be interesting to apply the same technique to the more broadly 
used gradient based hybrid MC algorithms~\citep{hybridmc} and other non-adaptive
accelerated methods (e.g. parallel tempering~\citep{paralleltempering}) where the alternative
techniques for determining convergence via diffusion approximations may be harder to apply.
More generally, the variational analysis could be a useful tool in making comparisons 
between the convergence properties of the latest MCMC algorithms without extensive numerical
simulation.
\begin{acknowledgments}
The authors wish to thank Cyrus Umrigar for discussions and the 
USDA-ARS plant pathogen systems biology group at Cornell University for 
computing resources. CRM acknowledges support from USDA-ARS project 
1907-21000-017-05. We also acknowledge support from NSF DMR 0705167. 
\end{acknowledgments}

\end{document}